\documentclass[universe,review,accept,pdftex]{Definitions/mdpi}
\usepackage{amsmath}
\usepackage{amsfonts}
\usepackage{amssymb,bm}
\usepackage{siunitx}
\usepackage{color}
\usepackage{braket}
\usepackage{graphics}
\firstpage{1}
\pubvolume{10}
\issuenum{2}
\articlenumber{84}
\pubyear{2024}
\copyrightyear{2024}
\externaleditor{Academic Editor: Panayiotis Stavrinos} 
\datereceived{22 January 2024}
\dateaccepted{7 February 2024}
\datepublished{9 February 2024}
\hreflink{https://doi.org/103390/\linebreak universe10020084} 

\Title{Prediction of the Expansion of the Universe made by Alexander Friedmann
and the Effect of Particle Creation in Cosmology }

\TitleCitation{Prediction of the Expansion of the Universe made by Alexander Friedmann
and the Effect of Particle Creation in Cosmology }

\Author{
Vladimir M. Mostepanenko \orcidB{}}

\AuthorNames{
 Vladimir M. Mostepanenko}

\AuthorCitation{ Mostepanenko, V.M.}

\address{%
Central Astronomical Observatory at Pulkovo of the Russian Academy of Sciences, Saint Petersburg, 196140, Russia\\
Peter the Great Saint Petersburg
Polytechnic University, Saint Petersburg, 195251, Russia\\
Kazan Federal University, Kazan, 420008, Russia}

\corres{Correspondence: vmostepa@gmail.com}

\abstract{This review devoted to the centenary of Alexander Friedmann's prediction
of the Universe expansion presents the results obtained by him in 1922
and 1924 and the sketch of their further developments. Special attention
is paid to the role of mathematics which enabled Friedmann to perform a
radical departure from the conventional practice of considering our
Universe as a static system. The effect of particle creation in expanding
Universe is discussed concurrently with the earlier investigated phenomenon
of pair creation from vacuum by external electric field. The numbers of
scalar and spinor particles created at different stages of the Universe
evolution are presented, and the possible role of the effect of creation
of particles in the formation of relativistic plasma and cold dark matter
after the inflationary period is noted. It is stressed that by introducing
the concept of expanding Universe Friedmann made a contribution
to understanding of the world around us compatible to those made by Ptolemy,
Copernicus, and Newton in previous epochs.}


\begin{document}
\section{Introduction}

A hundred years ago young mathematician Alexander Friedmann made an unexpected
prediction that our Universe expands with time. This prediction was in complete
contradiction with all the previous scientific concepts of the Universe
developed over the past millennia. One could mention the Ptolemy system, which
was geocentric, and the Copernicus, Kepler and Galilei system, which was
heliocentric. Based on the laws of mechanics and gravitation discovered by
him, Newton supposed \cite{1} that our Universe has an infinitely large volume,
contains infinitely many stars and exists in time forever. As a theologian,
Newton believed that the Universe is created by God. This means that not only
all material bodies, but also space and time are created in one creation act.
The question of whether the existence of the Universe in time is finite or
infinite must be solved by physics. All the mentioned pictures of the
world are static in the sense that they do not change with time. And even
Albert Einstein, after the creation of his general theory of relativity \cite{2}
especially modified its equations by introducing the cosmological constant in
order to obtain the static model of the Universe \cite{3} in agreement with
the concepts of all previous epochs.

Like Einstein, Friedmann described the Universe as a whole on a basis
of the general theory of relativity. In doing so, however, he restricted
himself to the minimum number of additional assumptions. Specifically, following
Newton and Einstein, he assumed that the 3-space of the Universe is homogeneous
and isotropic, i.e., there are no preferential points and preferential directions.
Otherwise, Friedmann acted as a mathematician by solving equations of
the fundamental general theory of relativity and looking for the results obtained
with no prejudice caused by some physical considerations like the desired static
character of any model of the Universe. Just this method of attack helped him
to make an outstanding prediction that our Universe expands with time which was
very soon confirmed by astronomical observations and became the cornerstone of
modern cosmology.

In this brief review, we discuss the scientific results by Alexander Friedmann
contained in his famous papers \cite{4,5} by making more emphasis on the
outstanding role of mathematics in their obtaining. According to
Friedmann's prediction, our Universe started its evolution from a point (the
so-called cosmological singularity), where it was characterized by the infinitely
large values of the scalar curvature and energy density. The Universe with a
3-space of negative or zero curvature expands infinitely long, whereas the
Universe with a 3-space of positive curvature expands to some maximum size and
then contracts down to a singular state.

Next, the outstanding phenomenon described by a unification of the general theory
of relativity and quantum field theory is considered. This is the effect of
particle-antiparticle pair creation from vacuum, which occurs due to the Universe
expansion, as was understood for the first time by Erwin Scr\"{o}dinger \cite{6}.
The effect of particle creation makes it possible to trace a mathematical
analogy between the well understood case of a nonstationary electric field and
the expanding space-time of the Friedmann Universe. Main approaches to the
definition of the concept of particles in the Friedmann cosmological models
and the calculation results for a creation rate are presented. The role of
the effect of particle creation at different stages of the Universe evolution,
including the epoch of inflation, is discussed.

This review is organized as follows. In Section 2, The Friedmann prediction of
the Universe expansion is considered with an emphasis on several facts of his
biography and mathematical educational background. Section 3 is devoted to the
effect of particle-antiparticle pair creation in the nonstationary electric
field. Section 4 contains the primary information relative to the effect of
particle creation in the Friedmann Universe. The crucial role of the effect of
particle creation in the transition period between the inflationary and the
radiation dominated stages of the Universe evolution is elucidated in Section 5.
A discussion of the fundamental importance of Friedmann's prediction for the
modern cosmology is presented in Section 6, and we will finish with the
conclusions in Section 7.

The system of units in which $c=\hbar=1$ is used, where $c$ is the speed of light and
$\hbar$ is the reduced Planck constant.

\section{Role of Mathematics in Friedmann's Prediction of the
Universe Expansion.}

It was difficult to imagine that Alexander Friedmann who was born on June 6,
1888 in the artistic family (father was a ballet artist and composer, mother
was a pianist \cite{7,8}) will become the outstanding mathematician and
physicist who will radically change our picture of the world. However, his
exceptional abilities in mathematics became apparent very early. In 1905,
while still a schoolboy, Alexander Friedmann, together with his schoolmate
Yakov Tamarkin, obtained new interesting results in the field of Bernoulli
numbers. In the next year, it was David Hilbert who recommended their paper
for publication in the prestigious mathematical Journal Mathematische
Annalen \cite{9}.

After the graduation from High School, Friedmann became a student
of the Department of Mathematics of the Saint Petersburg University where
he gained in-depth knowledge in different areas of mathematics and physics.
His successes were always evaluated as "excellent". Because of this, after a
graduation from the University in 1910, Friedmann was left at the
same Department for a preparation to the Professor position under a supervision
of the famous mathematician academician Vladimir Steklov. During the next
years, he published many papers containing solutions of several complicated
problems of mathematical physics. Starting from 1913, Friedmann
took an interest in the mathematical problems of dynamical meteorology,
aerodynamics, and hydrodynamics where he obtained a lot of fundamental
results which are well known to all experts in these fields and retain
their importance to the present day.

In 1920, Friedmann has had close contacts with several Professors of
the Petrograd (as St. Petersburg was called at that time) University who just
began delivering lectures in the recently developed quantum physics and
general theory of relativity. He has taken a great interest in the latter and
embarked upon giving lectures on tensor calculus at the University as an
introduction to the general theory of relativity. Friedmann was
inspired by the idea that the Universe around us is the Riemannian space-time
where all bodies move freely along the geodesic lines. This idea was
radically different from Newton's concept of the gravitational force which
acts between all material bodies through an empty space.

In 1922, Friedmann applied the formalism of the general theory of
relativity to the theoretical description of the Universe as a whole. As
mentioned in Section 1, he restricted himself to the minimum physical
assumptions by presuming that the 3-space of the Universe is homogeneous
and isotropic. In this regard, Friedmann followed Einstein \cite{3}
and de Sitter \cite{10}.

Mathematically, the requirement of homogeneity and isotropy of the 3-space
is expressed in the following distance (interval) $ds$ between two
infinitesimally close space-time points $x^i=(t,\chi,\theta,\varphi)$ and
$x^i+dx^i=(t+dt,\chi+d\chi,\theta+d\theta,\varphi+d\varphi)$:
\begin{linenomath}
\begin{equation}
ds^2=g_{ik}dx^idx^k\equiv dt^2-a^2(t)\left[d\chi^2+f^2(\chi)(d\theta^2+
\sin^2\theta d\varphi^2)\right],
\label{eq1}
\end{equation}
\end{linenomath}
where  $t$ is the time variable, and the spatial
coordinates $\chi, \theta$, and $\varphi$ are connected with the standard Cartesian
coordinates $(x^1,x^2,x^3)$ by the relations
\begin{equation}
x^1=a(t)f(\chi)\sin\theta\cos\varphi, \quad
x^2=a(t)f(\chi)\sin\theta\sin\varphi, \quad
x^3=a(t)f(\chi)\cos\theta.
\label{eq2}
\end{equation}

The quantity $a(t)$ in Eq.~(\ref{eq1}) has the dimension of length. It has the
meaning of the radius of curvature of space. As to the function $f(\chi)$,
it is defined as
\begin{linenomath}
\begin{equation}
f(\chi)=\left\{
\begin{array}{ll}
\sin\chi,&\kappa=1, \\
\sinh\chi,&\kappa=-1,\\
\chi,&\kappa=0,
\end{array}
\right.
\label{eq3}
\end{equation}
\end{linenomath}
where $\kappa$ is the sign of curvature of the 3-space ($\kappa=0$ corresponds
to the flat 3-space). Depending on the value of $f(\chi)$ in Eq.~(\ref{eq3}),
the interval (\ref{eq1}) relates to the closed space of the finite volume
$V=2\pi^2a^3(t)$ and positive curvature, to the open space of an infinite
volume and negative curvature, or to the quasi-Euclidean space of an infinite
volume and zero curvature.

Working as a mathematician, Friedmann solved the Einstein equations
\begin{linenomath}
\begin{equation}
R_{ik}-\frac{1}{2}g_{ik}R-\Lambda g_{ik}=8\pi GT_{ik}.
\label{eq4}
\end{equation}
\end{linenomath}
where $R_{ik}$ is the Ricci tensor describing the curvature of the space-time,
$R=g^{ik}R_{ik}$ is the scalar curvature, $\Lambda$ is the cosmological
constant, $G$ is the gravitational constant, $T_{ik}$ is the stress-energy
tensor of matter in the Universe, and $g_{ik}$ is the metrical tensor whose
components for $i,k=0,1,2,3$ are defined in Eq.~(\ref{eq1}) for the case of a
homogeneous and isotropic space. In this space, the stress-energy tensor is
diagonal, and its components have the meaning of the energy density,
$T_{0}^{\,0}=\varepsilon$, and pressure,
$T_{1}^{\,1}= T_{2}^{\,2}=T_{3}^{\,3}=-P$, of matter.
It is important that $R_{ik}$ and $R$ can be calculated for any given $g_{ik}$.
Note that the stress-energy tensor is also often called the energy-momentum tensor.

Substituting the metrical tensor $g_{ik}$ defined in Eq.~(\ref{eq1}) in
Eq.~(\ref{eq4}), one obtains two Friedmann equations for the unknown scale
factor $a(t)$ and the energy density $\varepsilon$
\begin{linenomath}
\begin{eqnarray}
&&
\frac{d^2a}{dt^2}=-\frac{4\pi G}{3}a(\varepsilon+3P)+\frac{1}{3}a\Lambda,
\nonumber \\
&&
\left(\frac{da}{dt}\right)^2=\frac{8\pi G}{3}a^2\varepsilon
-\kappa+\frac{1}{3}a^2\Lambda.
\label{eq5}
\end{eqnarray}
\end{linenomath}
We recall that the pressure $P$ is connected with the energy density by the
equation of state.

Note that initially Einstein introduced his equations (\ref{eq4}) with
$\Lambda=0$ \cite{2}. The cosmological term $\Lambda g_{ik}$ was introduced
by him later \cite{3} especially for obtaining the static model of the
Universe.

Friedmann considered the dust-like matter with the equation of
state $P=0$ (in our system of units $\varepsilon=\rho$ where $\rho$ is the
density of matter). The closed Universe $(\kappa=1)$ was considered by
Friedmann in Ref.~\cite{4} published in 1922 and the open Universe
$(\kappa=-1)$ --- in Ref.~\cite{5} published in 1924.

For instance, if $\kappa=1$ and $\Lambda=0$, Eq.~(\ref{eq5}) for the
dust-like matter is simplified to
\begin{eqnarray}
&&
\frac{d^2a}{dt^2}=-\frac{4\pi G}{3}a\rho,
\nonumber \\
&&
\left(\frac{da}{dt}\right)^2=\frac{8\pi G}{3}a^2\rho-1.
\label{eq6}
\end{eqnarray}

It is easy to check by the direct substitution that the solution of this
system of equations can be represented in the parametric form
\begin{linenomath}
\begin{eqnarray}
&&
a=\tilde{a}_0(1-\cos\eta),\qquad t=\tilde{a}_0(\eta-\sin\eta),
\nonumber \\
&&
\rho=\frac{3}{4\pi G}\,\frac{1}{{\tilde{a}}_0^2}\,\frac{1}{(1-\cos\eta)^3},
\qquad 0\leqslant\eta\leqslant 2\pi,
\label{eq7}
\end{eqnarray}
\end{linenomath}
where $\tilde{a}_0$ is the constant expressed via the total mass of matter
in the closed Universe $M$ as $\tilde{a}_0=2GM/(3\pi)$.

If one considers $t, \eta\ll 1$, Eq.~(\ref{eq7}) reduces to
\begin{linenomath}
\begin{equation}
a(t)\approx\left(\frac{9\tilde{a}_0}{2}\right)^{1/3}t^{2/3}, \qquad
\rho(t)\approx\frac{1}{6\pi Gt^2},
\label{eq8}
\end{equation}
\end{linenomath}
i.e., according to Friedmann, the evolution of the Universe
starts from a point-like state $a(0)=0$, where the density of matter
$\rho=\infty$.

The Universe expands with time until the maximum size $a_{\max}=2\tilde{a}_0$
reached at $\eta=\pi$, $t=\pi\tilde{a}_0$ and then contracts to a point
$a(2\pi\tilde{a}_0)=0$. For $\kappa=-1$ or 0 the expansion of the Universe
also starts from a point (called the cosmological singularity), where the
density of matter is infinitely large, but in this case the expansion goes
on infinitely long. Similar results were later obtained for the radiation
dominated Universe where matter has the equation of state $P=\varepsilon/3$
(see Ref.~\cite{11} for details). This equation of state describes the hot
Universe at the early stages of its evolution.

Thus, if $\Lambda=0$, all the solutions of Eq.~(\ref{eq5}) are nonstationary
and describe the expanding (or contracting in the case $\kappa=1$) Universe.
According to Friedmann, the static cosmological solution of
Einstein equations is possible only for the closed Universe ($\kappa=1$)
with the cosmological constant $\Lambda\neq 0$ satisfying the conditions
\begin{equation}
\Lambda=4\pi G(\varepsilon+3P), \qquad
4\pi Ga^2(\varepsilon+P)=1.
\label{eq9}
\end{equation}

Under these conditions Eq.~(\ref{eq5}) reduces to
\begin{linenomath}
\begin{equation}
\frac{d^2a}{dt^2}=\frac{da}{dt}=0,
\label{eq10}
\end{equation}
\end{linenomath}
which means that $a=const$. This is the static model of the Universe
obtained by Einstein \cite{3}. Friedmann did not discuss whether the Einstein
model is stable relative to some disturbance which occurs at a definite time.
This problem was investigated later after an experimental confirmation of the
Universe expansion (see Ref. \cite{11a} for a summary of the obtained results).
A more detailed consideration of the cosmological models with nonzero $\Lambda$
can be found in Ref.~\cite{12}.

We only mention the famous solution of Eq.~(\ref{eq5}) obtained by
de Sitter \cite{10} for the empty Universe with $\varepsilon=P=0$ but with
a nonzero cosmological constant $\Lambda$. In this case Eq.~(\ref{eq5})
takes the form
\begin{equation}
\frac{d^2a}{dt^2}=\frac{1}{3}a\Lambda, \qquad
\left(\frac{da}{dt}\right)^2=-\kappa+\frac{1}{3}a^2\Lambda.
\label{eq11}
\end{equation}

In the most simple, quasi-Euclidean case ($\kappa=0$), the De Sitter solution
of Eq.~(\ref{eq11}) is
\begin{linenomath}
\begin{equation}
a(t)=\tilde{a}_0\exp\left(\sqrt{\frac{\Lambda}{3}}t\right).
\label{eq12}
\end{equation}
\end{linenomath}
The closed ($\kappa=1$) and open ($\kappa=-1$) De Sitter solutions of
Eq.~(\ref{eq11}) are, respectively,
\begin{equation}
a(t)=\sqrt{\frac{3}{\Lambda}}\cosh\left(\sqrt{\frac{\Lambda}{3}}t\right),
\qquad
a(t)=\sqrt{\frac{3}{\Lambda}}\sinh\left(\sqrt{\frac{\Lambda}{3}}t\right).
\label{eq13}
\end{equation}

The scale factors in Eqs.~(\ref{eq12}) and (\ref{eq13}) are the exponentially
increasing with time functions which leave the scalar curvature constant,
$R=-4\Lambda$. The De Sitter solution found important applications in
theoretical description of the very early stages of the Universe evolution
near the cosmological singularity (see Section 5).

Although Friedmann's papers \cite{4,5} were published in the leading Journal
of that times, his remarkable results have not gained wide recognition over a
long period of time. Just after the publication of Friedmann's paper \cite{4},
Albert Einstein claimed \cite{13} that the solutions found by Friedmann do not
satisfy Eq.~(\ref{eq4}) of the general theory of relativity. It was, however,
Einstein who made a mistake in his note \cite{13}. After receiving of the
explanation letter from Friedmann, Einstein was obliged to recognize this
fact in another published note \cite{14}.

From the experimental viewpoint, the expansion of the Universe predicted by
Friedmann should manifest itself as a moving of all remote galaxies
away from the Earth. This would lead to the redshift of the light emitted by
them in accordance to the Doppler law. In fact the redshift of the light from
the Andromeda nebula was registered by Slipher \cite{15} as early as in 1913,
i.e., before the Friedmann prediction.

In a systematic way, the experimental law connecting the redshift in the spectra
of observable galaxies with the expansion of the Universe was found by Georges
Lema\^{i}tre in 1927 \cite{16} and Edwin Hubble in 1929 \cite{17} after they
identified the nebulas with remote galaxies \cite{18}. Lema\^{i}tre's paper
contains a rederivation of the main properties of expanding Universe from Einstein
equations with no citation of the papers \cite{4,5} by Friedmann who untimely passed
away of typhus on September 16, 1925 in the age of 37. Hubble's paper \cite{17}
does not cite Alexander Friedmann's papers as well. Later on the properties of
homogeneous isotropic metrics were studied by H.P. Robertson \cite{19} and
A.G. Walker \cite{20} whose papers also do not cite the Friedmann results.

In the meantime, after elaboration of the theory of a hot Universe by George
Gamov \cite{21}, the prediction of the relic radiation \cite{22} and its
discovery by Arno Penzias and Robert Wilson \cite{23}, it has become evident
that the Friedmann solution describing the expanding Universe formed the
foundation of modern cosmology and radically changed our picture of the
world. Starting from sixties of the last century, Friedmann's name as a
pioneer of the Universe expansion becomes more and more popular. The Friedmann
background as a mathematician played a crucial rile in his discovery which
was based on Einstein's equations of the general theory of relativity alone
with no unnecessary assumptions caused by either tradition or physical
intuition. This is one more example of what was characterized by E.P. Wigner
as "The unreasonable effectiveness of mathematics in the natural sciences"
\cite{24}.

Though being a mathematician, Friedmann considered his prediction
of the Universe expansion very seriously and expected that it will find the
experimental confirmation. In his book "The World as Space and Time"
written for a general reader and published in 1923 \cite{25}, Alexander
Friedmann not only explained the main concepts of Einstein's general theory
of relativity, but also discussed his own model of expanding Universe which
starts its evolution from a point. The front cover of this book is presented
in Figure 1. According to Friedmann's estimation contained in Ref.~\cite{25},
the interval between the Universe creation and the present day is of about
tens of billions of years (see the cover of the original publication of this book
in Figure 1). This estimation is in qualitative agreement with the modern
measurements which result in 13.7 billion years for the Universe age. Thus,
Friedmann predicted the most dramatic phenomenon of nature which
completely changed our picture of the world.

\begin{figure}[H]
\centerline{\hspace*{-2.7cm}
\includegraphics[width=2.in]{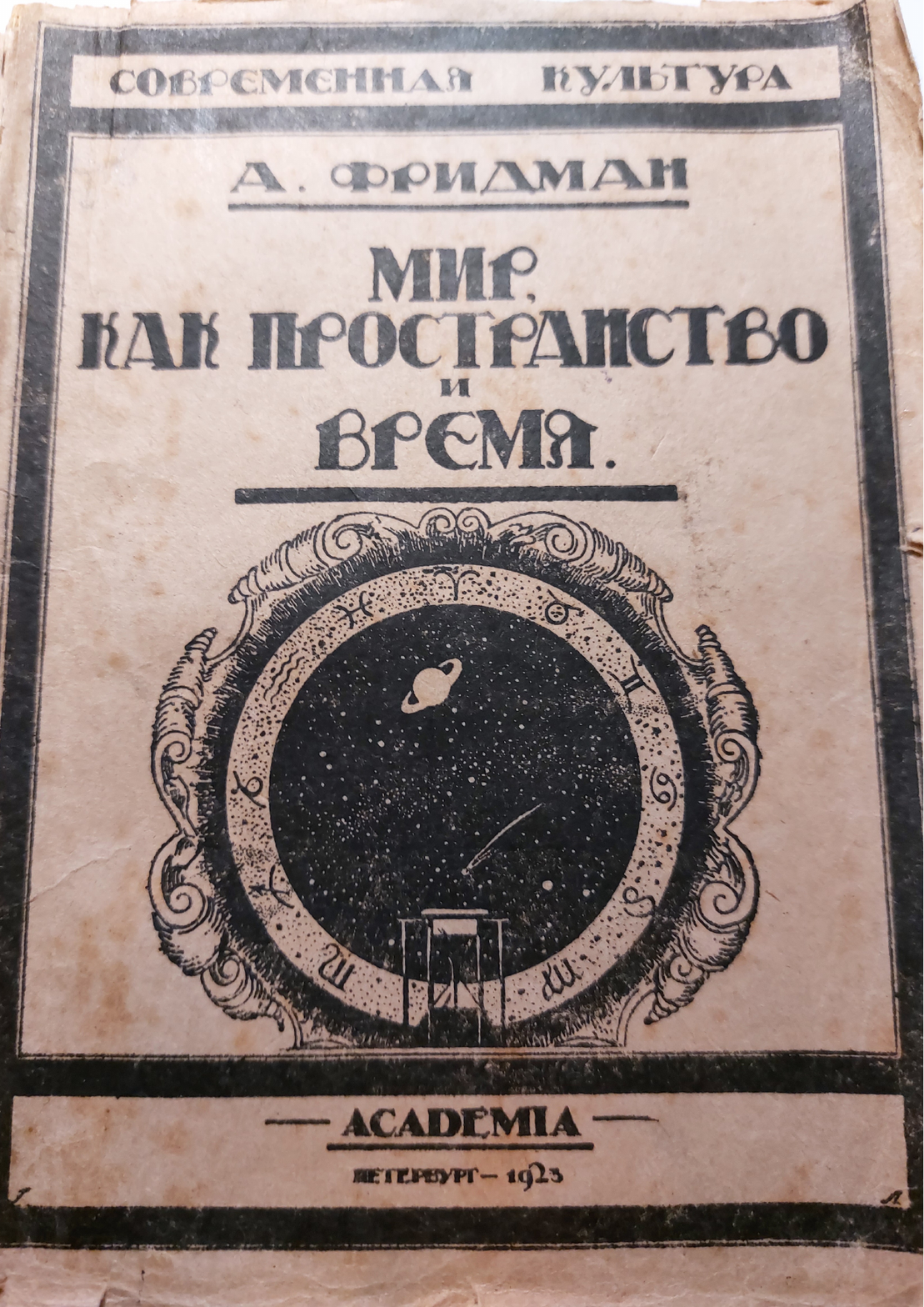}}
\caption{\label{fg1}
The front cover of the book \cite{25}. Translation: Modern culture.
A. Friedmann. The World as Space and Time. Academia, Petersburg - 1923.
}
\end{figure}

\section{Quantum Creation of Particle-Antiparticle Pairs in a Nonstationary
Electric Field}
\newcommand{\eb}{\mbox{\boldmath$E$}}
\newcommand{\pb}{\mbox{\boldmath$p$}}
\newcommand{\xb}{\mbox{\boldmath$x$}}
\newcommand{\asa}{\stackrel{\ast}{a}\vphantom{a}\!}
\newcommand{\bsa}{\stackrel{\ast}{b}\vphantom{b}\!}
\newcommand{\tp}{(\mbox{\boldmath$p$},t)}

As was mentioned in Section 1, the Universe expansion results in the effect
of particle creation from the vacuum state of quantized fields. This is the
quantum effect which is described by the quantum field theory in curved
space-time. It is most important at the very early stages of the Universe
evolution near the cosmological singularity where the Universe should be
considered as a quantum object.

The quantum field theory and the general theory of relativity are very
dissimilar theories. The former deals with the quantum fields defined
on a flat Minkowski space-time, whereas the latter treats the gravitational
field as a classical curved space-time. The quantum theory of gravitation
is not yet available in spite of numerous attempts to develop it undertaken
by many authors during half a century. It is possible, however, to consider
the quantized matter fields defined not on a Minkowski background, but on
a curved space-time of the expanding Universe. This theory is well elaborated
starting from the beginning of eighties of the last century (see, for
instance, the monographs \cite{26,27,28,29,30}).

Some basic concepts of quantum field theory in curved space-time, including
the concept of a particle, are, however, much more complicated and, unlike
the standard quantum field theory, and not defined uniquely.
Because of this, before considering the effect of particle creation
in the Friedmann Universe, we discuss in this section the creation of
particle-antiparticle pairs from vacuum by the nonstationary space
homogeneous electric field. Quantum electrodynamics allows to describe this
phenomenon in a quite transparent way \cite{31,32,33,34}. At the same time,
although conceptually the nonstationary electric field and the expanding
space-time of the Universe are quite different, mathematically the
description of the effect of particle creation in both cases turns out to
be very similar. Thus, the formalism briefly presented in this section will
provide rather useful guidance in the next section.

The spatially homogeneous nonstationary electric field directed along the
$z=x^3$ axis can be described by the vector potential
\begin{linenomath}
\begin{equation}
A^k(x)=\left(0,0,0,A^3(t)\right),
\label{eq14}
\end{equation}
\end{linenomath}
which leads to the field strength
\begin{linenomath}
\begin{equation}
\eb(t)=\left(0,0,-\frac{dA^3(t)}{dt}\right)=\left(0,0,E_z(t)\right),
\label{eq15}
\end{equation}
\end{linenomath}

It is assumed that the field is switched off at $t\rightarrow\pm\infty$, i.e.,
\begin{linenomath}
\begin{equation}
\lim_{t\to\pm\infty}A^3(t)=A_{\pm}^3=const, \qquad
\lim_{t\to\pm\infty}E_z(t)=0.
\label{eq16}
\end{equation}
\end{linenomath}

Let us consider first the complex scalar field of mass $m$ interacting with
the electric field (\ref{eq15}). A complete orthonormal set of solutions to
the Klein-Fock-Gordon equation
\begin{linenomath}
\begin{equation}
\left[\left(\frac{\partial}{\partial x^k}+ieA_k\right)
\left(\frac{\partial}{\partial x_k}+ieA^k\right)+m^2\right]\varphi(x)=0
\label{eq17}
\end{equation}
\end{linenomath}
in the case of a vector potential (\ref{eq14}) takes the form
\begin{linenomath}
\begin{equation}
\varphi_{\pb}^{\pm}(x)=\frac{1}{(2\pi)^{3/2}\sqrt{2\omega_{-}(\pb)}}\,
e^{i\pb\xb}g^{(\pm)}(\pb,t),
\label{eq18}
\end{equation}
\end{linenomath}
where ${\pb}=(p_1,p_2,p_3)$ is the momentum, the functions $g^{(\pm)}$ obey
the equation
\begin{linenomath}
\begin{equation}
\frac{d^2g^{(\pm)}(\pb,t)}{dt^2}+\omega^2(\pb,t)g^{(\pm)}(\pb,t)=0,
\qquad
\omega^2(\pb,t)=m^2+p_{\bot}^2+\left(p_3-eA_2(t)\right)^2
\label{eq19}
\end{equation}
\end{linenomath}
and the following notations are used
\begin{linenomath}
\begin{equation}
p_{\bot}^2=p_1^2+p_2^2, \qquad
\omega_{-}(\pb)=\lim_{t\to-\infty}\omega(\pb,t).
\label{eq20}
\end{equation}
\end{linenomath}

Equation (\ref{eq19}) is the equation of oscillator with a variable frequency
\cite{33,34}. The positive- and negative-frequency solutions of this equation
are defined by the following asymptotic behavior:
\begin{linenomath}
\begin{equation}
\lim_{t\to-\infty}g^{(\pm)}(\pb,t)=e^{\pm i\omega_{-}(\small\pb)\,t}.
\label{eq21}
\end{equation}
\end{linenomath}

An operator of the complex scalar field is
\begin{linenomath}
\begin{equation}
\varphi(x)=\int d^3p\left[\varphi_{\pb}^{(-)}(x)a_{\pb}^{(-)}+
\varphi_{-\pb}^{(+)}(x)a_{\pb}^{(+)}\right],
\label{eq22}
\end{equation}
\end{linenomath}
where $a_{\pb}^{(-)}$ is the annihilation operator for particles and
$a_{\pb}^{(+)}$ is the creation operator for antiparticles defined at
$t\rightarrow -\infty$ when the scalar field is free. The vacuum state
at $t\rightarrow -\infty$ is defined as
\begin{linenomath}
\begin{equation}
a_{\pb}^{(-)}|0_{\rm in}\rangle=\asa_{\pb}^{(-)}|0_{\rm in}\rangle=0,
\label{eq23}
\end{equation}
\end{linenomath}
where $\asa_{\pb}^{(-)}$ is the annihilation operator for antiparticles
(the creation operator for particles is notated as $\asa_{\pb}^{(+)}$).

The Hamiltonian of the complex scalar field is defined by \cite{35}
\begin{eqnarray}
&&
H^{(0)}(t)=\int d^3xT_{00}(x)=\int d^3x\left[
\vphantom{\left(\frac{\partial}{\partial x^k}\right)}
2\partial_0\varphi^{\ast}(x)\partial_0\varphi(x)\right.
\nonumber \\
&&~~~~
-\left.\left(\frac{\partial}{\partial x^k}-ieA_k\right)\varphi^{\ast}(x)
\left(\frac{\partial}{\partial x_k}+ieA^k\right)\varphi(x)
+m^2\varphi^{\ast}(x)\varphi(x)\right].
\label{eq24}
\end{eqnarray}

Substituting Eq.~(\ref{eq22}) in Eq.~(\ref{eq24}) and performing the
integration with respect to $\xb$ and to one of the momenta using
(\ref{eq18}), one obtains
\begin{linenomath}
\begin{eqnarray}
&&
H^{(0)}(t)=\int d^3p\omega(\pb,t)\left[E(\pb,t)\left(\asa_{\pb}^{(+)}a_{\pb}^{(-)}
+\asa_{-\pb}^{(-)}a_{-\pb}^{(+)}\right)\right.
\nonumber \\
&&~~~~~
+\left.\vphantom{\left(\asa_{\pb}^{(+)}a_{\pb}^{(-)}\right)}
F(\pb,t)\asa_{\pb}^{(+)}a_{-\pb}^{(+)}+
F^{\ast}(\pb,t)\asa_{-\pb}^{(-)}a_{\pb}^{(-)}\right],
\label{eq25}
\end{eqnarray}
\end{linenomath}
where
\begin{linenomath}
\begin{eqnarray}
&&
E(\pb,t)=\frac{1}{2\omega_{-}(\pb)\omega(\pb,t)}\left[
\left|\frac{dg^{(+)}(\pb,t)}{dt}\right|^2+\omega^2(\pb,t)
\left|g^{(+)}(\pb,t)\right|^2\right],
\nonumber \\
&&
F(\pb,t)=\frac{1}{2\omega_{-}(\pb)\omega(\pb,t)}\left[
\left(\frac{dg^{(+)}(\pb,t)}{dt}\right)^2+\omega^2(\pb,t)
{g^{(+)}}^2(\pb,t)\right],
\nonumber \\
&&
E^2(\pb,t)-|F(\pb,t)|^2=1.
\label{eq26}
\end{eqnarray}
\end{linenomath}

Using Eq.~(\ref{eq21}), Eq.~(\ref{eq26}) leads to
\begin{equation}
\lim_{t\to -\infty}E(\pb,t)=1, \qquad \lim_{t\to -\infty}F(\pb,t)=0.
\label{eq27}
\end{equation}

As a result, at $t\rightarrow -\infty$ The Hamiltonian (\ref{eq24})
takes the diagonal form
\begin{linenomath}
\begin{equation}
\lim_{t\to -\infty}H^{(0)}(t)=
\int d^3p\omega_{-}(\pb)\left(\asa_{\pb}^{(+)}a_{\pb}^{(-)}
+\asa_{-\pb}^{(-)}a_{-\pb}^{(+)}\right),
\label{eq28}
\end{equation}
\end{linenomath}
as it should be for the Hamiltonian of free field.

At any $t$, in the presence of a nonstationary electric field, the
Hamiltonian (\ref{eq25}) can be diagonalized by means of the canonical
Bogoliubov transformations which preserve the commutation relations for
the creation-annihilation operators
\begin{linenomath}
\begin{eqnarray}
&&
a_{\pb}^{(-)}=\alpha_{\pb}^{\ast}(t)b_{\pb}^{(-)}(t)-
\beta_{\pb}(t)b_{-\pb}^{(+)}(t),
\nonumber \\
&&
\asa_{\pb}^{(-)}=\alpha_{-\pb}^{\ast}(t)\bsa_{\pb}^{(-)}(t)-
\beta_{-\pb}(t)\bsa_{-\pb}^{(+)}(t),
\label{eq29}
\end{eqnarray}
\end{linenomath}
where
\begin{linenomath}
\begin{equation}
|\alpha_{\pb}(t)|^2-|\beta_{\pb}(t)|^2=1.
\label{eq30}
\end{equation}
\end{linenomath}
Note that an addition of the creation operators to the annihilation ones
in Eq.~(\ref{eq29}) due to the action of a nonstationary external field is
equivalent to the fact that the negative-frequency solution of the wave
equation defined at $t\rightarrow -\infty$ becomes the linear combination
of the negative- and positive-frequency solutions defined at a later
time $t$.

If the coefficients $\alpha_{\pb}(t)$ and $\beta_{\pb}(t)$ are given
by
\begin{linenomath}
\begin{equation}
\frac{\beta_{\pb}(t)}{\alpha_{\pb}(t)}=\frac{E\tp-1}{F^{\ast}\tp}, \qquad
|\beta_{\pb}(t)|^2=\frac{1}{2}[E\tp -1],
\label{eq31}
\end{equation}
\end{linenomath}
the Hamiltonian (\ref{eq25}) takes a diagonal form at any $t$ \cite{33}
\begin{linenomath}
\begin{equation}
H^{(0)}(t)=
\int d^3p\omega(\pb,t)\left[\bsa_{\pb}^{(+)}(t)b_{\pb}^{(-)}(t)
+\bsa_{-\pb}^{(-)}(t)b_{-\pb}^{(+)}(t)\right].
\label{eq32}
\end{equation}
\end{linenomath}
In doing so the operators $\bsa_{\!\pb}^{\!(+)}(t)$ and
${b}_{\pb}^{(-)}(t)$ can be considered as the creation and annihilation
operators of quasiparticles defined at the moment $t$. The quasiparticle
vacuum is defined by
\begin{linenomath}
\begin{equation}
b_{\pb}^{(-)}(t)|0_t\rangle=\bsa_{\!\pb}^{(-)}(t)|0_t\rangle=0.
\label{eq33}
\end{equation}
\end{linenomath}

It is easily seen that
\begin{linenomath}
\begin{equation}
\lim_{t\to -\infty}\beta_{\pb}(t)=0, \qquad \lim_{t\to -\infty}\alpha_{\pb}(t)=1,
\label{eq34}
\end{equation}
\end{linenomath}
so that the creation and annihilation operators of quasiparticles at
$t\rightarrow -\infty$ coincide with the creation and annihilation operators
$a_{\pb}^{(\pm)}$, $\asa_{\pb}^{(\pm)}$ and the quasiparticle vacuum
$\vert 0_{-\infty}\rangle$=$\vert 0_{\rm in}\rangle$ defined in Eq.~(\ref{eq23}).

Now one can find the number of scalar quasiparticles with the momentum
$\pb$ and antiparticles with the momentum $-\pb$ created from the vacuum
state $\vert 0_{\rm in}\rangle$
\begin{linenomath}
\begin{equation}
N_{\pb}^{(0)}(t)=\langle 0_{\rm in}|\bsa_{\pb}^{(+)}(t)b_{\pb}^{(-)}(t)|0_{\rm in}\rangle
=\langle 0_{\rm in}|b_{-\pb}^{(+)}(t)\bsa_{-\pb}^{(-)}(t)|0_{\rm in}\rangle
=|\beta_{\pb}(t)|^2\delta^{3}(\pb=0).
\label{eq35}
\end{equation}
\end{linenomath}
These quasiparticles pairs were created by the electric field during the time
interval from $-\infty$ to $t$ in the space of an infinitely large volume $V$.
Taking into account that
\begin{linenomath}
\begin{equation}
\delta^{3}(\pb=0)=\frac{1}{(2\pi)^3}\int d^3x=\frac{V}{(2\pi)^3},
\label{eq36}
\end{equation}
\end{linenomath}
for the total number of scalar quasiparticle pairs with any momentum created
in the unit space volume one obtains
\begin{linenomath}
\begin{equation}
n^{(0)}(t)=\frac{1}{V}\int d^3pN_{\pb}^{(0)}(t)=
\frac{1}{(2\pi)^3}\int d^3p|\beta_{\pb}(t)|^2.
\label{eq37}
\end{equation}
\end{linenomath}

In the asymptotic limit $t\rightarrow \infty$, the electric field is switched
off and in this "out" region the quasiparticles described by the operators
$b_{\pb}^{(\pm)}(\infty)$, $\bsa_{\!\pb}^{\!(\pm)}(\infty)$ become
the real free particles. Thus, the total number of real boson pairs created by
the electric field during the time of its existence is
\begin{linenomath}
\begin{equation}
n^{(0)}=\lim_{t\to\infty}n^{(0)}(t)=
\frac{1}{(2\pi)^3}\int d^3p[\lim_{t\to\infty}|\beta_{\pb}(t)|^2].
\label{eq38}
\end{equation}
\end{linenomath}

Similar results have been obtained for the fields and particles with nonzero
spin. By omitting the technical details, here we present only several facts
concerning the case of spinor particles. Thus, after the separation of
variables in Dirac equation written for the spinor field interacting
with the space homogeneous nonstationary electric field (\ref{eq14}),
(\ref{eq15}), it reduces to the oscillator equation with the complex
frequency \cite{33,34}
\begin{linenomath}
\begin{equation}
\frac{d^2f^{(\pm)}\tp}{dt^2}+\left[\omega^2\tp+ie\frac{dA_3(t)}{dt}\right]
f^{(\pm)}\tp=0,
\label{eq39}
\end{equation}
\end{linenomath}
where $\omega^2(\pb,t)$ is presented in Eq.~(\ref{eq19}) and the positive-
and negative-frequency solutions are defined by the following asymptotic
behaviors:
\begin{linenomath}
\begin{equation}
\lim_{t\to-\infty}f^{(\pm)}\tp=
\frac{1}{\sqrt{4\omega_{-}(\pb)[\omega_{-}(\pb)+p^3-eA_{-}^3]}}\,
e^{\pm i\omega_{-}(\pb)t}.
\label{eq40}
\end{equation}
\end{linenomath}

The Hamiltonian of spinor field interacting with the electric field
(\ref{eq14}) is given by
\begin{linenomath}
\begin{eqnarray}
&&
H^{(1/2)}(t)=\sum_{r=1,2}
\int d^3p\omega(\pb,t)\left[E(\pb,t)\left(\asa_{\pb r}^{(+)}a_{\pb r}^{(-)}
-\asa_{-\pb r}^{(-)}a_{-\pb r}^{(+)}\right)\right.
\nonumber \\
&&~~~~~
+\left.\vphantom{\left(\asa_{\pb r}^{(+)}a_{\pb r}^{(-)}\right)}
F(\pb,t)\asa_{\pb r}^{(+)}a_{-\pb r}^{(+)}+
F^{\ast}(\pb,t)\asa_{-\pb r}^{(-)}a_{\pb r}^{(-)}\right],
\label{eq41}
\end{eqnarray}
\end{linenomath}
where the index $r=1,2$ corresponds to two possible spin projections on the
axis $x^3$ and the coefficients $E$ and $F$ are defined as
\begin{linenomath}
\begin{eqnarray}
&&
E(\pb,t)=\frac{4(m^2+p_{\bot}^2)}{\omega\tp}\,{\rm Im}\left[
{f^{(+)}}^{\ast}\tp\frac{{df^{(+)}}\tp}{dt}\right]-
\frac{p_3-eA_3}{\omega\tp},
\nonumber \\
&&
E^2(\pb,t)+|F(\pb,t)|^2=1.
\label{eq42}
\end{eqnarray}
\end{linenomath}

Similar to the case of a scalar field, the Hamiltonian (\ref{eq41}) becomes
diagonal in the asymptotic limit $t\to-\infty$
\begin{linenomath}
\begin{equation}
H^{(1/2)}(t)=\sum_{r=1,2}
\int d^3p\omega_{-}(\pb)\left[\asa_{\pb r}^{(+)}a_{\pb r}^{(-)}
-\asa_{-\pb r}^{(-)}a_{-\pb r}^{(+)}\right].
\label{eq34}
\end{equation}
\end{linenomath}

At any $t$ the Hamiltonian (\ref{eq41}) can be
diagonalized by the canonical Bogoliubov transformation preserving the
anticommutation relations between the creation and annihilation operators
of spinor particles
\begin{linenomath}
\begin{eqnarray}
&&
a_{\pb r}^{(-)}=\alpha_{\pb}^{\ast}(t)b_{\pb r}^{(-)}(t)-
\beta_{\pb}(t)b_{-\pb r}^{(+)}(t),
\nonumber \\
&&
\asa_{\pb r}^{(-)}=\alpha_{-\pb}^{\ast}(t)\bsa_{\pb r}^{(-)}(t)-
\beta_{-\pb}(t)\bsa_{-\pb r}^{(+)}(t),
\label{eq44}
\end{eqnarray}
\end{linenomath}
where
\begin{linenomath}
\begin{equation}
|\alpha_{\pb}(t)|^2+|\beta_{\pb}(t)|^2=1.
\label{eq45}
\end{equation}
\end{linenomath}

If the coefficients of the Bogoliubov transformation (\ref{eq44}) are
equal to
\begin{linenomath}
\begin{equation}
\frac{\beta_{\pb}(t)}{\alpha_{\pb}(t)}=\frac{1-E\tp}{F^{\ast}\tp}, \qquad
|\beta_{\pb}(t)|^2=\frac{1}{2}[1-E\tp],
\label{eq46}
\end{equation}
\end{linenomath}
the Hamiltonian (\ref{eq41}) takes the diagonal form at any $t$ in terms of
the creation and annihilation operators of quasiparticles \cite{33}
\begin{linenomath}
\begin{equation}
H^{(1/2)}(t)=\sum_{r=1,2}
\int d^3p\omega(\pb,t)\left[\bsa_{\pb r}^{(+)}(t)b_{\pb r}^{(-)}(t)-
\bsa_{-\pb r}^{(-)}(t)b_{-\pb r}^{(+)}(t)\right].
\label{eq47}
\end{equation}
\end{linenomath}

Similar to Eq.~(\ref{eq33}), the vacuum state of quasiparticles is defined as
\begin{linenomath}
\begin{equation}
b_{\pb r}^{(-)}(t)|0_t\rangle=\bsa_{\pb r}^{(-)}(t)|0_t\rangle=0.
\label{eq48}
\end{equation}
\end{linenomath}

The number of spinor quasiparticles with momentum $\pb$ and spin projection
$r$ (and respective antiquasiparticles) created from the ground state
$\vert 0_{in}\rangle$ during the time interval from $-\infty$ to $t$ is
given by
\begin{linenomath}
\begin{equation}
N_{\pb r}^{(1/2)}(t)=\langle 0_{\rm in}|\bsa_{\pb r}^{(+)}(t)b_{\pb r}^{(-)}(t)
|0_{\rm in}\rangle
=\langle 0_{\rm in}|b_{-\pb r}^{(+)}(t)\bsa_{-\pb r}^{(-)}(t)|0_{\rm in}\rangle
=|\beta_{\pb}(t)|^2\delta^{3}(\pb=0).
\label{eq49}
\end{equation}
\end{linenomath}
This number does not depend on the spin state $r$.

The total number of fermion quasiparticle pairs created in the unit space volume
during the time interval from $-\infty$ to $t$ is obtained from Eq.~(\ref{eq49})
with the help of Eq.~(\ref{eq36})
\begin{linenomath}
\begin{equation}
n^{(1/2)}(t)=\frac{1}{V}\sum_{r=1,\,2}\int d^3pN_{\pb r}^{(1/2)}(t)=
\frac{2}{(2\pi)^3}\int d^3p|\beta_{\pb}(t)|^2.
\label{eq50}
\end{equation}
\end{linenomath}

Thus, the total number of real fermion pairs created in the unit volume by the
electric field is
\begin{linenomath}
\begin{equation}
n^{(1/2)}=
\frac{2}{(2\pi)^3}\int d^3p[\lim_{t\to\infty}|\beta_{\pb}(t)|^2].
\label{eq51}
\end{equation}
\end{linenomath}

The most simple exactly solvable example allowing an exact calculation of the
numbers of created pairs (\ref{eq38}) and (\ref{eq51}) is the electric field
of the form \cite{31,33}
\begin{linenomath}
\begin{equation}
A^3=-\frac{E_0}{k_0}\tanh(k_0t), \qquad
E_z(t)=\frac{E_0}{\cosh^2(k_0t)}.
\label{eq52}
\end{equation}
\end{linenomath}
This field is switched off in the asymptotic regimes $t\rightarrow\pm\infty$
(see Figure 2).

\begin{adjustwidth}{-\extralength}{0cm}
\begin{figure}[H]
\vspace*{-12cm}
\centerline{\hspace*{-2.7cm}
\includegraphics[width=8.5in]{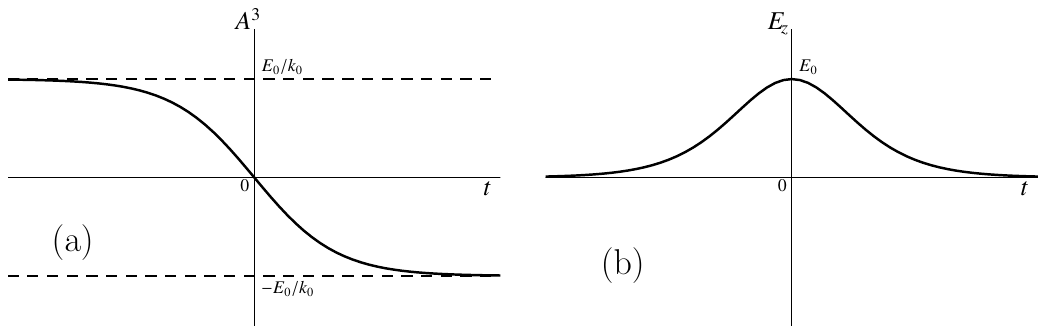}}
\vspace*{-12.cm}
\caption{\label{fg2}
 (a) The component of the vector potential and (b) the strength of
the space homogeneous nonstationary electric field (\ref{eq52}), which is
switched off in the asymptotic regimes $t\rightarrow \pm\infty$, are shown
as the functions of time.
}
\end{figure}
\end{adjustwidth}

In the limiting case $k_0\rightarrow 0$, Eq.~(\ref{eq52}) describes the space
homogeneous constant electric field. Thus, Eqs.~(\ref{eq38}) and (\ref{eq51})
allow rederivation of the famous Schwinger results for the pair creation
from vacuum by strong constant field derived by him \cite{36,37} using another
formalism.

For the inflationary cosmology (see Section 5), the effect of exponential
growth of the number of scalar particles created with some values of momentum
by the periodic in time external field is of much importance. This effect
was independently discovered in Ref.~\cite{38} for the sinusoidally
depending on time $A_3$ and in Ref.~\cite{39} for the electric field of
arbitrary form with a period $T$
\begin{linenomath}
\begin{equation}
A_3(t+T)=A_3(t)
\label{eq53}
\end{equation}
\end{linenomath}
during the interval $[0, nT]$. Outside of this interval, the electric field was
assumed to be equal to zero, so that $A_{3-}=A_{3+}=const$.

It was shown that the number of pairs of scalar particles created by the
periodic field with some momenta $\pb$ belonging to the instability zones of
oscillator equation during the time $nT$ is the exponentially increasing
function of the number of field periods $n$ \cite{39}
\begin{linenomath}
\begin{equation}
n_{\pb}^{(0)}=|\beta_{\pb}(nT)|^2=\frac{\sinh^2[nD(\pb)]}{\sinh^2D(\pb)}\,
\frac{1}{4\omega_{+}^2(\pb)}\left[\omega_{+}^2(\pb) g_1(\pb,T)+
\left.\frac{dg_2(\pb,t)}{dt}\right|_{t=T}\right]^2.
\label{eq54}
\end{equation}
\end{linenomath}
Here, $g_1(\pb,t)$ and $g_2(\pb,t)$ are the solutions of the
oscillator equation (\ref{eq19}) satisfying the initial conditions
\begin{linenomath}
\begin{equation}
\left\{
\begin{array}{ll}
g_1(\pb,0)=0,&{\ }\\[1mm]
\left.\frac{g_1(\pb,\,t)}{dt}\right|_{t=0}=1,&{\ }
\end{array}
\right.
\qquad
\left\{
\begin{array}{ll}
g_2(\pb,0)=1,&{\ }\\[1mm]
\left.\frac{g_2(\pb,\,t)}{dt}\right|_{t=0}=0,&{\ }
\end{array}
\right.
\label{eq55}
\end{equation}
\end{linenomath}
and $coshD(\pb)=g_2(\bf {p},T)$, $\omega_{+}(\pb)=\omega_{-}(\pb)$.

From Eq.~(\ref{eq54}) it is seen that the number of created pairs
$n_{\pb}^{(0)}$ increases with the number of field periods $n$ as
$exp[2nD(\pb)]$.

The effect of particle creation from vacuum by an electric field is not
observed yet because it becomes sizable for the fields of the order of
$m^2/e$ which are too large ($\sim 10^{16}$~V/cm for electrons). With a
discovery of graphene, where the fermion quasiparticles are massless or
very light, the possibility was proposed to observe the creation of this
quasiparticles in much weaker fields \cite{39a,39b,39c,39d,39e,39f}.
This is some kind of a condensed matter analogy to Schwinger's particle
creation from vacuum in quantum electrodynamics.

The above brief discussion allows to conclude that in quantum field theory
and, specifically, in quantum electrodynamics, a description of the
effect of particle creation from vacuum by external field is based on the
S-matrix picture. The concept of real particles is defined in the "in" and
"out" regions where the external electric field is switched off. It is
common knowledge that in the absence of external fields the theory is
invariant relative to the transformations from the Poincar\'{e} group
whose Casimir operators classify particles by the values of their mass
and spin \cite{35}. Thus, in curved space-time, which does not become
flat in the asymptotic regions, one could expect difficulties with the
definition of the concept of particles. In the next section it is shown
that in the case of expanding Universe these difficulties can be solved
in close analogy to the concept of quasiparticles in the presence of a
nonstationary electric field.

\section{The Effect of Particle Creation in the Friedmann Universe}
\newcommand{\xv}{\mbox{\boldmath$x$}}

As discussed in Section 2, the Friedmann models of the Universe are
described by the interval (\ref{eq19}). By solving the Einstein equations
(\ref{eq4}) for the metrical tensor $g_{ik}$ defined in Eq.~(\ref{eq1}),
one obtains the scale factors of the closed, open, and quasi-Euclidean
models. The matter fields (scalar and spinor, for instance) should be
considered on the background of curved space-time defined in Eq.~(\ref{eq1}).

The general covariant generalization of the Klein-Fock-Gordon equation
(\ref{eq17}) with the electric field $A_k=0$ is given by
\begin{linenomath}
\begin{equation}
\left(\nabla_k\nabla^k+\xi R+m^2\right)\varphi(x)=0,
\label{eq56}
\end{equation}
\end{linenomath}
where $\nabla_k$ is the covariant derivative and $\xi$ is the so-called
coupling coefficient. The most simple case $\xi=0$ is referred to as the
minimal coupling. In the case $\xi=1/6$ considered in Refs.~\cite{40,41},
Eq.~(\ref{eq56}) becomes invariant under the conformal transformations when
$m=0$. This is called the conformal coupling.

As was first noticed by Schr\"{o}dinger \cite{6}, the positive-frequency
solution of Eq.~(\ref{eq56}) with $\xi=0$ in the space-time with metric
(\ref{eq1}) defined at some moment $t_0$ becomes the linear combination
of the negative- and positive-frequency solutions of the same equation
defined at a later moment $t$. Schr\"{o}dinger interpreted this fact as
a creation of matter merely by the expansion of the Universe.

In more detail, the theory of particle creation in the expanding Universe
was considered by Parker \cite{42,43} (see also the review \cite{44}).
This consideration was restricted to the quasi-Euclidean model with a
flat 3-space ($\kappa=0$). The space-time of this model, as well as of
the other Friedmann models, is not asymptotically flat. Therefore, as
discussed in the end of Section 3, the standard concept of particles
used in quantum field theory is not applicable.

To solve this problem, Parker elaborated the concept of the so-called
adiabatic particles. For this purpose, the solution of Eq.~(\ref{eq56})
was searched in the form of WKB-like approximation including some
unknown function, which was next determined from the demand that the
number of created particles and of its derivatives of several first
orders would take the minimum values. The creation rate of scalar
particles defined in this way in the present epoch of the Universe
evolution was calculated and found to be negligibly small. Similar
approach was applied to the effect of creation of spinor particles in
the expanding Universe with $\kappa=0$ \cite{45}. A simple model was
proposed where the scalar particles described by the field equation
with minimal coupling are created near the cosmological singularity
with a black-body spectrum \cite{46}.

The separation of variables in Eq.~(\ref{eq56}) for the quasi-Euclidean,
closed and open models of the Universe [$\kappa=0,\,\pm 1$ in
Eqs.~(\ref{eq1}) and (\ref{eq3})] was made in the form
\begin{linenomath}
\begin{equation}
\varphi_J(x)=\frac{1}{a(\eta)}g_{\lambda}(\eta)\Phi_J(\xv),
\label{eq57}
\end{equation}
\end{linenomath}
where the dimensionless time variable $\eta$ is connected with the proper
synchronous time $t$ by $dt=a(\eta)d\eta$, $\lambda$ is the dimensionless
momentum quantum number connected with the magnitude of the physical
momentum by $p=\lambda/a(\eta)$, $J=(\lambda,l,m)$ is the collective index,
and the explicit expressions for the functions $\varphi_J$ in terms of the
associated Legendre polinomials and spherical harmonics $Y_{lm}$ in spaces
with $\kappa=0,\,\pm 1$ were found in Refs.~\cite{47,48,49}.

Substitution of Eq.~(\ref{eq57}) in Eqs.~(\ref{eq1}) and (\ref{eq17})
results in the following equation for the functions $g_{\lambda}$:
\begin{linenomath}
\begin{equation}
\frac{d^2g_{\lambda}(\eta)}{d\eta^2}+\left[\omega_{\lambda}^2(\eta)-
q(\eta)\right]g_{\lambda}(\eta)=0,
\label{eq58}
\end{equation}
\end{linenomath}
where
\begin{linenomath}
\begin{equation}
\omega_{\lambda}^2(\eta)=\lambda^2+m^2a^2(\eta),\qquad
q(\eta)=6\left(\frac{1}{6}-\xi\right)\left[\frac{1}{a(\eta)}\,
\frac{d^2a(\eta)}{d\eta^2}+\kappa\right].
\label{eq59}
\end{equation}
\end{linenomath}
For $\kappa=0,\,-1$ the dimensionless momentum $\lambda$ varies from 0 to
$\infty$ and for $\kappa=1$ it holds $\lambda=1,\,2,\,3,\,\ldots\,$.

It is seen that Eq.~(\ref{eq58}) describes the oscillator with a variable
frequency like it was for a scalar field interacting with the nonstationary
electric field [compare with Eq.~(\ref{eq19})]. In this case the role of
electric field is played by the time-dependent scale factor of the Universe.
Equation (\ref{eq58}) takes the most simple form for the scalar field with
conformal coupling $(\xi=1/6)$ which is physically the most natural
generalization of the Klein-Fock-Gordon equation in curved space-time
\cite{40,41}. The point is that the massless particles are not characterized
by the parameter with a dimension of length and, thus, the corresponding
field equation must be invariant with respect to the conformal
transformations. Because of this, we consider Eqs.~(\ref{eq56}) and
(\ref{eq58}) with $\xi=1/6$. As a result, the function $g_{\lambda}$
satisfies the equation
\begin{linenomath}
\begin{equation}
\frac{d^2g_{\lambda}(\eta)}{d\eta^2}+\omega_{\lambda}^2(\eta)g_{\lambda}(\eta)=0,
\label{eq60}
\end{equation}
\end{linenomath}
where
\begin{linenomath}
\begin{equation}
\omega_{\lambda}^2(\eta)=\lambda^2+m^2a^2(\eta).
\label{eq61}
\end{equation}
\end{linenomath}

An important difference of the scale factor of expanding Universe $a(\eta)$
from the vector potential of an external field $A_3(t)$ is that $a(\eta)$
does not become constant at any $\eta$ which means that the space-time of
the Universe does not become static. In this situation, the corpuscular
interpretation of the field can be performed at some moment $\eta_0$ by imposing
the initial conditions on the solutions of Eq.~(\ref{eq60})
\begin{linenomath}
\begin{equation}
g_{\lambda}(\eta_0)=\frac{1}{\sqrt{\omega_{\lambda}(\eta_0)}},
\qquad
\left.\frac{dg_{\lambda}(\eta)}{d\eta}\right|_{\eta=\eta_0}=i
\omega_{\lambda}(\eta_0)g_{\lambda}(\eta_0)
\label{eq62}
\end{equation}
\end{linenomath}
and defining the positive- and negative-frequency solutions of
Eq.~(\ref{eq56}) as
\begin{linenomath}
\begin{eqnarray}
&&
\varphi_J^{(+)}(x)=\frac{1}{\sqrt{2}\,a(\eta)}\,g_{\lambda}(\eta)\Phi_J^{\ast}(\xv),
\nonumber \\
&&
\varphi_J^{(-)}(x)=\frac{1}{\sqrt{2}\,a(\eta)}\,g_{\lambda}^{\ast}(\eta)\Phi_J(\xv).
\label{eq63}
\end{eqnarray}
\end{linenomath}

Similar to the case of the nonstationary electric field, the functions
$\varphi_J^{(\pm)}$ lose the meaning of the negative- and positive-frequency
solutions at a later moment $\eta>\eta_0$.

The field operator of the complex scalar field is defined similar to
Eq.~(\ref{eq22})
\begin{linenomath}
\begin{equation}
\varphi(x)=\int\! d\mu(J)\left[\varphi_J^{(-)}(x)a_J^{(-)}+
\varphi_J^{(+)}(x)a_J^{(+)}\right],
\label{eq64}
\end{equation}
\end{linenomath}
where the measure on the set of quantum numbers is different for different
values of $\kappa$
\begin{linenomath}
\begin{equation}
\int\! d\mu(J)=\left\{
\begin{array}{ll}
\int\limits_0^{\infty}d\lambda\sum\limits_{l=0}^{\infty}\sum\limits_{m=-l}^{l},
&{\ }\kappa=-1,\,0,\\[5mm]
\sum\limits_{\lambda=1}^{\infty}\sum\limits_{l=0}^{\lambda-1}\sum\limits_{m=-l}^{l},
&{\ }\kappa=1.
\end{array}\right.
\label{eq65}
\end{equation}
\end{linenomath}
Then the vacuum state at the moment $\eta_0$ is defined as
\begin{linenomath}
\begin{equation}
a_J^{(-)}|0_{\eta_0}\rangle=\,\,\asa_J^{(-)}|0_{\eta_0}\rangle=0.
\label{eq66}
\end{equation}
\end{linenomath}

From the above it becomes clear that it is not possible to introduce the
universal concept of particles in the expanding space-time of the Friedmann
Universe. It is possible, however, to define the quasiparticles depending
on time like it was done in Section 3 for the case of a nonstationary
electric field using the method of diagonalization of the Hamiltonian of
quantized field. Such an approach was suggested in Refs.~\cite{50,51} as an
alternative to the adiabatic particles introduced in Refs. \cite{42,43,44}.

It is important, however, that the stress-energy tensor and respective
Hamiltonian of the quantized scalar field satisfying Eq.~(\ref{eq56}) with
$\xi=1/6$ should be obtained by the variation of the action not with
respect to the field $\varphi$ but with respect to the metrical tensor $g^{ik}$.
This is the so-called metrical stress-energy tensor \cite{52}. As a result,
the metrical Hamiltonian of the scalar field in the space-time of expanding
Universe takes the form
\begin{linenomath}
\begin{eqnarray}
&&
H^{(0)}(\eta)=a^2(\eta)\int\!d^3xf^2(\chi)\sin\theta T_{00}^{\rm metr}(x)
\nonumber \\
&&~~
=a^2(\eta)\int\!d^3xf^2(\chi)\sin\theta\left[2
\frac{\partial\varphi^{\ast}(x)}{\partial\eta}\,\frac{\partial\varphi(x)}{\partial\eta}
-a^2(\eta)g^{ik}
\frac{\partial\varphi^{\ast}(x)}{\partial x^i}\,\frac{\partial\varphi(x)}{\partial x^k}
\right.
\label{eq67} \\
&&~~~~\left.
+a^2(\eta)\left(m^2+\frac{1}{6}R\right)\varphi^{\ast}(x)\varphi(x) -\frac{1}{3}
\left(R_{00}+\nabla_0\nabla_0-a^2(\eta)\nabla_k\nabla^k\right)
\varphi^{\ast}(x)\varphi(x)
\vphantom{\frac{\partial\varphi^{\ast}(x)}{\partial\eta}}\right].
\nonumber
\end{eqnarray}
\end{linenomath}

After a substitution of Eq.~(\ref{eq64}) in Eq.~(\ref{eq67}), using the
properties of functions (\ref{eq62}), one obtains \cite{53,54}
\begin{linenomath}
\begin{eqnarray}
&&
H^{(0)}(\eta)=\int\!d\mu(J)\omega_{\lambda}(\eta)\left[
E_J(\eta)\left(\asa_J^{(+)}a_J^{(-)}+\asa_{\bar{J}}^{(-)}a_{\bar{J}}^{(+)}\right)\right.
\nonumber \\
&&~~~\left.
+F_J(\eta)\asa_J^{(+)}a_{\bar{J}}^{(+)}+F_J^{\ast}(\eta)\asa_{\bar{J}}^{(-)}a_J^{(-)}\right],
\label{eq68}
\end{eqnarray}
\end{linenomath}
where $\bar{J}=(\lambda,l,-m)$ and the coefficients $E_J$ and $F_J$ are
expressed via the solutions of Eq.~(\ref{eq60}) as
\begin{linenomath}
\begin{eqnarray}
&&
E_J(\eta)=\frac{1}{2\omega_{\lambda}(\eta)}
\left(\left|\frac{dg_{\lambda}(\eta)}{d\eta}\right|^2
+\omega_{\lambda}^2(\eta)|g_{\lambda}(\eta)|^2\right),
\nonumber \\
&&
F_J(\eta)=\frac{(-1)^m}{2\omega_{\lambda}(\eta)}
\left[\left(\frac{dg_{\lambda}(\eta)}{d\eta}\right)^2
+\omega_{\lambda}^2(\eta)g_{\lambda}^2(\eta)\right],
\nonumber \\
&&
E_J^2(\eta)-|F_J(\eta)|^2=1.
\label{eq69}
\end{eqnarray}
\end{linenomath}

From Eq.~(\ref{eq69}) it is seen that $E_J(\eta)$ in fact depends on
$\lambda$ and does not depend on $l$ and $m$, whereas $F_J(\eta)$ depends
also on $m$. The quantity $E_J(\eta)$ has the meaning of the adiabatic
invariant of the oscillator (\ref{eq60}), (\ref{eq61}). From Eq.~(\ref{eq62})
it follows
\begin{linenomath}
\begin{equation}
E_J(\eta_0)=1,\qquad F_J(\eta_0)=0,
\label{eq70}
\end{equation}
\end{linenomath}
i.e., the Hamiltonian (\ref{eq68}) takes the diagonal form at the initial
moment $\eta_0$
\begin{linenomath}
\begin{equation}
H^{(0)}(\eta_0)=\int\!d\mu(J)\omega_{\lambda}(\eta_0)
\left(\asa_J^{(+)}a_J^{(-)}+\asa_{\bar{J}}^{(-)}a_{\bar{J}}^{(+)}\right)
\label{eq71}
\end{equation}
\end{linenomath}
in perfect analogy to Eq.~(\ref{eq28}) obtained for the case of electric field.

Similar to the case of a nonstationary electric field, at any moment the
Hamiltonian (\ref{eq68}) can be diagonalized by the canonical Bogoliubov
transformations
\begin{linenomath}
\begin{eqnarray}
&&
a_J^{(-)}=\alpha_J^{\ast}(\eta)b_J^{(-)}(\eta)-
(-1)^m\beta_J(\eta)b_{\bar{J}}^{(+)}(\eta),
\nonumber \\
&&
\asa_J^{(-)}=\alpha_J^{\ast}(\eta)\bsa_J^{(-)}(\eta)-(-1)^m\beta_J(\eta)
\bsa_{\bar{J}}^{(+)}(\eta),
\nonumber \\
&&
|\alpha_J(\eta)|^2-|\beta_J(\eta)|^2=1.
\label{eq72}
\end{eqnarray}
\end{linenomath}

For this purpose, the coefficients $\alpha_J$ and $\beta_J$ should be
chosen as
\begin{linenomath}
\begin{equation}
\frac{\beta_J(\eta)}{\alpha_J(\eta)}=(-1)^m\frac{E_J(\eta)-1}{F_J^{\ast}(\eta)},
\qquad
|\beta_J(\eta)|^2=\frac{1}{2}[E_J(\eta)-1].
\label{eq73}
\end{equation}
\end{linenomath}
Substituting Eq.~(\ref{eq72}) with the coefficients (\ref{eq73}) in
Eq.~(\ref{eq68}), one finds that the Hamiltonian of the scalar in the
Friedmann Universe takes the diagonal form
\begin{linenomath}
\begin{equation}
H^{(0)}(\eta)=\int\!d\mu(J)\omega_{\lambda}(\eta)
\left(\bsa_J^{(+)}b_J^{(-)}+\bsa_J^{(-)}b_J^{(+)}\right)
\label{eq74}
\end{equation}
\end{linenomath}
at any moment $\eta$.

The annihilation operators for quasiparticles and antiquasiparticles
give the possibility to define the time-dependent vacuum state by the
equation
\begin{linenomath}
\begin{equation}
b_J^{(-)}(\eta)|0_{\eta}\rangle=\bsa_J^{(-)}(\eta)|0_{\eta}\rangle=0,
\label{eq75}
\end{equation}
\end{linenomath}
which is similar to Eq.~(\ref{eq33}) in the case of a nonstationary
electric field. It is evident that
\begin{linenomath}
\begin{equation}
b_J^{(\pm)}(\eta_0)=a_J^{(\pm)}, \qquad
\bsa_J^{(\pm)}(\eta_0)=\asa_J^{(\pm)}.
\label{eq76}
\end{equation}
\end{linenomath}

Next one can define the number of quasiparticle pairs created from the
vacuum state $\vert 0_{\eta_0}\rangle$ during the time interval from
$\eta_0$ to $\eta$ in the unit space volume
\begin{linenomath}
\begin{eqnarray}
&&
n^{(0)}(\eta)=\frac{1}{2\pi^2a^3(\eta)}\int\!d\mu(\lambda)
\langle 0_{\eta_0}|\bsa_J^{(+)}(\eta)b_J^{(-)}(\eta)|0_{\eta_0}\rangle
\nonumber \\
&&~~~
=\frac{1}{2\pi^2a^3(\eta)}\int\!d\mu(\lambda)
\langle 0_{\eta_0}|b_{\bar{J}}^{(+)}(\eta)\bsa_{\bar{J}}^{(-)}(\eta)|0_{\eta_0}\rangle
\nonumber \\
&&~~~
=\frac{1}{2\pi^2a^3(\eta)}\int\!d\mu(\lambda)|\beta_J(\eta)|^2,
\label{eq77}
\end{eqnarray}
\end{linenomath}
where
\begin{linenomath}
\begin{equation}
\int\!d\mu(\lambda)=\left\{
\begin{array}{ll}
\int\limits_0^{\infty}\lambda^2d\lambda, &{\ }\kappa=-1,\,0, \\[4mm]
\sum\limits_{\lambda=1}^{\infty}\lambda^2, &{\ }\kappa=1.
\end{array}\right.
\label{eq78}
\end{equation}
\end{linenomath}

For calculation of the number of created scalar particles it is reasonable
to put $\eta_0=0$ and impose on the scale factor $a(\eta)$ the requirement
of smoothness at the initial moment $\eta_0=0$. This requirement does not
contradict to the fact that at the point $\eta=0$ there was the cosmological
singularity where the invariants of the curvature tensor become infinitely
large.

The typical scale factors used in the Friedmann cosmological models have
the form $a(t)=a_0t^q$, see, for instance, Eq.~(\ref{eq8}) where $q=2/3$
for the dust-like matter $\varepsilon=\rho, P=0$. In the vicinity of the
cosmological singularity matter is in the radiation dominated state
$(P=\varepsilon/3)$. In this case $q=1/2$.

Calculations show that in the epoch $t\ll m^{-1}$ the number of
quasiparticle pairs (\ref{eq77}) created in the unit volume does not depend
on the value of $q$ \cite{53,55}
\begin{linenomath}
\begin{equation}
n^{(0)}(t)=\frac{m^3}{24\pi^2}.
\label{eq79}
\end{equation}
\end{linenomath}
An independence of the result (\ref{eq79}) on time means that the decrease
of the quasiparticle density due to the Universe expansion is compensated
by the creation of new quasiparticles.

In the epoch $t\gg m^{-1}$, for the radiation dominated equation of state
$(q=1/2)$ one obtains \cite{55}
\begin{linenomath}
\begin{equation}
n^{(0)}(t)=5.3\times 10^{-4}m^3(mt)^{-3/2}.
\label{eq80}
\end{equation}
\end{linenomath}
It was shown \cite{55} that for $t\gg m^{-1}$ similar result
\begin{linenomath}
\begin{equation}
n^{(0)}(t)\sim m^3(mt)^{-3q}
\label{eq81}
\end{equation}
\end{linenomath}
holds for any $q$ satisfying the inequalities $0<q<2/3$.

The corresponding results have been obtained also for the energy density
of created pairs (see Refs.~\cite{53,55} and the review \cite{54}).

The creation of spinor particles in the space-time of expanding Universe
can be considered in perfect analogy with the scalar case although the
mathematical formalism becomes more involved. Thus, the general covariant
generalization of the Dirac equation takes the form
\begin{linenomath}
\begin{equation}
\left(i\gamma^k(x){\overrightarrow \nabla_k}-m\right)\psi(x)=0,
\label{eq82}
\end{equation}
\end{linenomath}
where $\overrightarrow \nabla_k$ is the covariant derivative of a bispinor
$\psi$ in the Riemannian space-time and $\gamma^k(x)$ is the 4-vector
relative to the general coordinate transformations, which is expressed via
the standard Dirac $\gamma$-matrices and the tetrad $h_{(a)}^{\,\,\, k}$ as
\begin{linenomath}
\begin{equation}
\gamma^k(x)=h_{(a)}^{\,\,\, k}\gamma^a.
\label{eq83}
\end{equation}
\end{linenomath}

An important characteristic feature of Eq.~(\ref{eq82}) is that in the
limiting case $m\rightarrow 0$ it becomes invariant under the conformal
transformations with no additional modifications.

The separation of variables in Eq.~(\ref{eq82}) for the space-time
(\ref{eq1}) was performed in Refs.~\cite{48,56}. It results in the
oscillator equation for the time-dependent functions $f_{\lambda\pm}$
\begin{linenomath}
\begin{equation}
\frac{d^2f_{\lambda\pm}(\eta)}{d\eta^2}+\left[\omega_{\lambda}^2(\eta)\pm
im\frac{da(\eta)}{d\eta}\right]f_{\lambda\pm}(\eta)=0,
\label{eq84}
\end{equation}
\end{linenomath}
where $\omega_{\lambda}$ is defined in Eq.~(\ref{eq61}). It is seen that
although physically the space-time of expanding Universe has little in
common with the nonstationary electric field considered in Section 3,
mathematically Eq.~(\ref{eq84}) is similar to Eq.~(\ref{eq39}). In doing
so, the mass of a spinor field in Eq.~(\ref{eq84}) plays the same role as
the electric charge in Eq.~(\ref{eq39}), whereas the scale factor of the
Universe $a$ is akin the vector potential $A_3$.

The positive- and negative-frequency solutions of Eq.~(\ref{eq84}) at the
moment $\eta_0$ are defined by the following initial conditions \cite{54}
\begin{linenomath}
\begin{equation}
f_{\lambda\pm}^{(+)}(\eta_0)=\pm\left[
\frac{\omega_{\lambda}(\eta_0)\mp ma(\eta_0)}{\omega_{\lambda}(\eta_0)}\right]^{1/2},
\qquad
f_{\lambda\pm}^{(-)}(\eta_0)=\left[
\frac{\omega_{\lambda}(\eta_0)\pm ma(\eta_0)}{\omega_{\lambda}(\eta_0)}\right]^{1/2}.
\label{eq85}
\end{equation}
\end{linenomath}

It holds also
\begin{linenomath}
\begin{equation}
\left.\frac{df_{\lambda\pm}^{(+)}(\eta)}{d\eta}\right|_{\eta=\eta_0}\!\!\!\!=
i\omega_{\lambda}(\eta_0)f_{\lambda\pm}^{(+)}(\eta_0),
\qquad
\left.\frac{df_{\lambda\pm}^{(-)}(\eta)}{d\eta}\right|_{\eta=\eta_0}\!\!\!\!=
-i\omega_{\lambda}(\eta_0)f_{\lambda\pm}^{(-)}(\eta_0).
\label{eq86}
\end{equation}
\end{linenomath}

Now the operator of the spinor field can be written in the form
\begin{linenomath}
\begin{equation}
\psi(x)=\int\!d\mu(J)\left[\psi_J^{(-)}(x)a_J^{(-)}+
\psi_J^{(+)}(x)a_J^{(+)}\right],
\label{eq87}
\end{equation}
\end{linenomath}
where the collective index $J$ includes four quantum numbers
$J=(\lambda,j,l,M)$. In the case $\kappa=0,\,-1$ it holds $0\le\lambda<\infty$,
$j=1/2,\,3/2,\,\ldots\,$, for $\kappa=1$ one has $\lambda=3/2,\,5/2,\,\ldots\,$,
$j=1/2,\,3/2,\,\ldots\,,\lambda-1$,
and in both cases $l=j\pm 1/2$, $-j\leqslant M\leqslant j$.

The vacuum state at the moment $\eta_0$ is defned by Eq.~(\ref{eq66}).
Substituting Eq.~(\ref{eq87}) in the Hamiltonian of the spinor field
\begin{linenomath}
\begin{equation}
H^{(1/2)}(\eta)=\frac{i}{2}a^3(\eta)\int\!d^3xf^2(\chi)\,sin\theta\,
\psi^{+}(x){\stackrel{\leftrightarrow}{\partial}}_\eta\psi(x),
\label{eq88}
\end{equation}
\end{linenomath}
one obtains it in the same form as in Eq.~(\ref{eq24}), but with the
coefficients $E_J$ and $F_J$ expressed via the solutions of Eq.~(\ref{eq84})
\begin{linenomath}
\begin{eqnarray}
&&
E_J(\eta)=\frac{1}{\omega_{\lambda}(\eta)}\left[ma(\eta)\left(1-
|f_{\lambda+}^{(+)}|^2\right)-\lambda{\rm Re}\,\left(f_{\lambda-}^{(-)}
f_{\lambda-}^{(+)}\right)\right],
\nonumber\\
&&
F_J(\eta)=\frac{1}{\omega_{\lambda}(\eta)}\left[ma(\eta)f_{\lambda+}^{(+)}
f_{\lambda-}^{(+)}\
-\frac{\lambda}{2}\left({f_{\lambda+}^{(+)}}^2-
{f_{\lambda-}^{(+)}}^2\right)\right],
\nonumber \\
&&
E_J^2(\eta)+|F_J(\eta)|^2=1.
\label{eq89}
\end{eqnarray}
\end{linenomath}

These coefficients satisfy the initial conditions (\ref{eq70}). As a
consequence, at the moment $\eta_0$ the Hamiltonian $H^{(1/2)}(\eta_0)$
takes the diagonal form. At any moment the Hamiltonian of the spinor
field can be diagonalized by the Bogoliubov transformations
\begin{linenomath}
\begin{eqnarray}
&&
a_J^{(-)}=\alpha_J^{\ast}(\eta)b_J^{(-)}(\eta)-\beta_J(\eta)b_{\bar{J}}^{(+)}(\eta),
\nonumber \\
&&
\asa_J^{(-)}=\alpha_J^{\ast}(\eta)\bsa_J^{(-)}(\eta)+\beta_J(\eta)
\bsa_{\bar{J}}^{(+)}(\eta),
\nonumber \\
&&
|\alpha_J(\eta)|^2+|\beta_J(\eta)|^2=1.
\label{eq90}
\end{eqnarray}
\end{linenomath}
which preserve the anticommutation relations for the creation and
annihilation operators.

The Hamiltonian $H^{(1/2)}(\eta)$ takes the diagonal form (\ref{eq74}) at
any moment $\eta$ if the Bogoliubov coefficients are defined as
\begin{linenomath}
\begin{equation}
\frac{\beta_J(\eta)}{\alpha_J(\eta)}=\frac{1-E_J(\eta)}{F_J^{\ast}(\eta)},
\qquad
|\beta_J(\eta)|^2=\frac{1}{2}[1-E_J(\eta)].
\label{eq91}
\end{equation}
\end{linenomath}

The number of spinor quasiparticle pairs created in the unit space volume
is given by \cite{56}
\begin{linenomath}
\begin{eqnarray}
&&
n^{(1/2)}(\eta)=\frac{1}{\pi^2a^3(\eta)}\int\!d\mu(\lambda)
\langle 0_{\eta_0}|\bsa_J^{(+)}(\eta)b_J^{(-)}(\eta)|0_{\eta_0}\rangle
\nonumber \\
&&~~~
=\frac{1}{\pi^2a^3(\eta)}\int\!d\mu(\lambda)
\langle 0_{\eta_0}|b_{\bar{J}}^{(+)}(\eta)\bsa_{\bar{J}}^{(-)}(\eta)|0_{\eta_0}\rangle
\nonumber \\
&&~~~
=\frac{1}{\pi^2a^3(\eta)}\int\!d\mu(\lambda)|\beta_J(\eta)|^2,
\label{eq92}
\end{eqnarray}
\end{linenomath}
where $J=(\lambda,j,j\pm1/2,M)$, $\bar{J}=(\lambda,j,j\mp 1/2,-M)$ and
\begin{linenomath}
\begin{equation}
\int\!d\mu(\lambda)=\left\{
\begin{array}{ll}
\int\limits_0^{\infty}\!d\lambda\left(\lambda^2-\frac{\kappa}{4}
\vphantom{\frac{1}{4}}\right),
&{\ }\kappa=-1,\,0, \\[4mm]
\sum\limits_{\lambda=3/2}^{\infty}\left(\lambda^2-\frac{1}{4}\right),
&{\ }\kappa=1.
\end{array}\right.
\label{eq93}
\end{equation}
\end{linenomath}

It is notable that that the geometric nature of the spinor field reveals
itself by the presence of $\kappa$ in the measure of integration (\ref{eq93}).

By using Eqs.~(\ref{eq89}), (\ref{eq91}), and (\ref{eq92}), one can calculate
the number of spinor quasiparticles created at different epochs of the
Universe evolution for the scale factors of power type $a(t)=a_0t^q$. Thus,
for $t\ll m^{-1}$ in the case $\kappa=0$ it holds \cite{57}
\begin{linenomath}
\begin{equation}
n^{(1/2)}(t)=\frac{q^2}{3q-1}\,\frac{m^2}{t}.
\label{eq94}
\end{equation}
\end{linenomath}

From the comparison of Eqs.~(\ref{eq79}) and (\ref{eq94}) it is seen that
$n^{(1/2)}/n^{(0)}\sim (mt)^{-1}\gg 1$. The additional terms which appear for
$\kappa=\pm 1$ are much smaller than (\ref{eq94}).

For the epoch $t\gg m^{-1}$ the density of created spinor quasiparticles
is expressed as \cite{54}
\begin{linenomath}
\begin{equation}
n^{(1/2)}(t)=K^{(q)}m^3(mt)^{-3q}+\frac{3q^2m}{256\pi t^2},
\label{eq95}
\end{equation}
\end{linenomath}
where for the radiation dominated matter $(q=1/2)$ the coefficient is equal
to $K^{(1/2)}=3.9\times 10^{-3}$.

As is seen from the above, in the nonstationary curved space-time the concept
of particle loses its unique meaning. The effect of particle creation takes
place with any concept of a particle but, for instance, the number of created
adiabatic particles may differ from the number of quasiparticles defined by
the method of diagonalization of Hamiltonian by means of the Bogoliubov
transformations. The covariant quantity describing the quantum effects in
the nonstationary space-time of cosmological models is the renormalized
vacuum expectation value of the stress-energy tensor of quantized fields.
This quantity includes the contributions of both the particle creation and
vacuum polarization (see the monographs \cite{26,27,28,29,30} for the
obtained results).

\section{The Role of Particle Creation in the Transition from Inflationary to
Radiation Dominated Epochs and Further Developments}

As discussed in Sections 2 and 4, at the radiation dominated stage of its
evolution the Friedmann Universe is described by the power-type scale
factor $a(t)=a_0t^{1/2}$. This result is obtained by solving the classical
Einstein equations, and it does not take into account the quantum effects.
However, the extension of the radiation dominated scale factor down to the
Planck time $t_{Pl}=G^{1/2}=5.39\times 10^{-44}~$s creates serious problems. One
of them is the following. Calculation shows that at the Planck time the size
of the Universe was $a(t_{Pl})\sim10^{-3}~$cm, i.e., it was by almost 30 orders
of magnitude larger than the Planck length $l_{pl}=1.62\times 10^{-33}~$cm
traveled by light during $t_{Pl}$.

From this it follows that if the radiation scale factor were valid down to
$t_0=0$, at $t=t_{Pl}$ the Universe would comprised of about $10^{89}$
causally disconnected domains. No evidence, however, was found regarding
differences in the temperature of relic radiation received from different
directions in the sky. Thus, the initial expansion of the Universe had
happened much faster than it is predicted by the power-type law. This
inconsistence was called the horizon problem.

As noted in the end of Section 4, the covariant description of the vacuum
quantum effects in curved space-time is provided by the renormalized
vacuum expectation value of the stress-energy tensor of quantized matter
fields. In Refs.~\cite{58,59} published in the beginning of 1980, this
quantity was considered as a single source of curved space-time of the
Universe. For this purpose, the self-consistent Einstein equations with no
cosmological term
\begin{linenomath}
\begin{equation}
R_{ik}-\frac{1}{2}Rg_{ik}=8\pi G\langle 0|T_{ik}|0\rangle_{\rm ren}
\label{eq96}
\end{equation}
\end{linenomath}
have been solved and the De Sitter solutions were obtained. For instance,
for a stress-energy tensor of massless scalar field in the closed Friedmann
model the solution of Eq.~(\ref{eq96}) is
\begin{linenomath}
\begin{equation}
a(t)=\sqrt{\frac{G}{360\pi}}\,\cosh\left(t\sqrt{\frac{360\pi}{G}}\right),
\label{eq97}
\end{equation}
\end{linenomath}
i.e., for $t>t_{Pl}$ the Universe expansion goes on exponentially fast. The
comparison of Eq.~(\ref{eq97}) with Eq.~(\ref{eq13}) shows that the vacuum
stress-energy tensor of quantized scalar field plays the same role as the
cosmological term in Einstein's equations (\ref{eq4}) with $T_{ik}=0$. In
Ref.~\cite{59} it was shown that under the impact of creation of scalarons
and their subsequent decay into the standard particles the exponentially
fast De Sitter expansion of the Universe passes into the power-type
expansion of the radiation dominated stage of its evolution.

In 1981 another approach to the understanding of exponentially fast
expansion of the Universe near the cosmological singularity was suggested
which was called inflation \cite{60}. This approach introduces the minimally
coupled classical scalar field $\phi=\phi(t)$ called the inflaton field
with the Lagrangian
\begin{linenomath}
\begin{equation}
L(\phi)=\frac{1}{2}\left[\left(\frac{d\phi}{dt}\right)^2-m^2\phi^2\right].
\label{eq98}
\end{equation}
\end{linenomath}

The corresponding Klein-Fock-Gordon equation in the space-time with metric
(\ref{eq1}) is
\begin{linenomath}
\begin{equation}
\frac{d^2\phi}{dt^2}+\frac{3}{a}\,\frac{da}{dt}\,\frac{d\phi}{dt}+m^2\phi=0.
\label{eq99}
\end{equation}
\end{linenomath}

In the simplest case of the quasi-Euclidean model $(\kappa=0)$, the second
equality in Eq.~(\ref{eq5}) with $\Lambda=0$ is
\begin{linenomath}
\begin{equation}
\frac{1}{a^2}\left(\frac{da}{dt}\right)^2=\frac{8\pi G}{3}\varepsilon,
\label{eq100}
\end{equation}
\end{linenomath}
where the space-time is determined by the energy density of the inflaton
field
\begin{linenomath}
\begin{equation}
\varepsilon=\frac{1}{2}\left[\left(\frac{d\phi}{dt}\right)^2+m^2\phi^2\right].
\label{eq101}
\end{equation}
\end{linenomath}

According to Ref.~\cite{61}, at the inflationary stage, the second term on
the left-hand side of Eq.~(\ref{eq99}) is much larger than the first one
and the term $m^2\phi^2$ in Eq.~(\ref{eq101}) is much larger than
$(d\phi/dt)^2$. As a result, the scale factor $a(t)$ found from
Eqs.~(\ref{eq100}) and (\ref{eq101}) takes the quasi exponential form
\begin{linenomath}
\begin{equation}
a(t)=\tilde{a}_0 \,\exp\left(2m\sqrt{\frac{\pi G}{3}}\phi t\right).
\label{eq102}
\end{equation}
\end{linenomath}

In succeeding years, many papers were published devoted to different
versions of the inflationary cosmology (see, e.g., Refs.~\cite{62,63,64,65}
and the monographs \cite{66,67}).

The model of inflation has inspired a renewed interest in the effect of
particle creation in the nonstationary external fields and in the space-time
of expanding Universe. The point is that in the end of inflationary stage
of the Universe evolution the energy density becomes very low and the
inflaton field oscillates near the minimum of its potential [in Eq.~(\ref{eq98})
the simplest potential $V=m^2\phi^2/2$ is chosen]. The standard elementary
particles were created during this period, which was called the process of
reheating after inflation \cite{66,68}.

The theory of the process of reheating is based on the effect of exponential
growth of the number of particle-antiparticle pairs created from vacuum by
the time-periodic field with some momenta belonging to the instability zones
of Klein-Fock-Gordon equation (see Section 3). In this case the role of a
periodic electric field is played by the oscillating inflaton field \cite{69,70}.
The theory of reheating after inflation was elaborated by many authors (see,
e.g., Refs.~\cite{71,72,73,74,75,76,77,78,79,80}). The main features of this theory
are summarized in Ref.~\cite{61}.

During the last 25 years the effect of particle creation in expanding Universe
continued to attract considerable attention of experts in quantum field theory
and cosmology. Here we mention only several papers devoted to this subject.
Thus, in Ref.~\cite{81} the effect of creation of light particles called moduli
during and after inflation was investigated not only numerically but also
analytically. It was shown that the dominant contribution to the particle
creation is given by the long-wavelength fluctuations of light scalar fields
generated during inflation.

In Ref.~\cite{82} the complex WKB approximation technique was used to study the
thermal particle creation in both the black holes and in the space-time of
expanding Universe. According to the results obtained, the temperature of the
particle spectrum is determined by the slope of scale factor of the
cosmological model.

The effect of particle creation in the anisotropic expanding Universe
(see the pioneer Ref.~\cite{83}) was further considered in Ref.~\cite{84}
using the formalism of squeezed vacuum states for a minimally coupled
scalar field. The semiclassical Einstein equations of the form of
Eq.~(\ref{eq96}), but in the anisotropic case, were discussed. Note that
Ref.~\cite{83} presented the powerful regularization method for the vacuum
stress-energy tensor and derived the dynamical equations for the nonstationary
Bogoliubov coefficients which were actively used in both anisotropic and
isotropic spaces.

In Ref.~\cite{85} the above Eqs.~(\ref{eq81}) and (\ref{eq95}) were used
to describe the creation of superheavy scalar and spinor particles whose
decay could explain the baryon number of the Universe and the nature of
cold dark matter. Note that previously the creation of superheavy particles
as the constituents of dark matter in various models of inflation was
analyzed in Ref.~\cite{86}. It was hypothesized that the decay products of
superheavy constituents of cold dark matter are observed as the cosmic rays
of ultra-high energy \cite{85}.

The method of diagonalization of Hamiltonian of quantized massless scalar
field with minimal coupling was used in Ref.~\cite{87} to calculate the
particle creation rate in the expanding Universe of quasi-Euclidean type.
It was assumed that the background matter is described by the equation
of state of a perfect fluid which may violate the strong energy condition
$\varepsilon+P\geqslant 0, \varepsilon+3P\geqslant 0$.
According to the results obtained,
the particle creation rate decreases with time if the strong energy condition
is satisfied and increases otherwise.

The creation of dark matter particles, which interact only gravitationally,
in expanding Universe of the quasi-Euclidean type was investigated in
Ref.~\cite{88}. In the suggested model, the real scalar field with an
arbitrary coupling $\xi$, whose quanta can be considered the candidates for
dark matter particles, enters into the Lagrangian density along with the
inflaton field, but does not interact with it. By calculating the particle
creation rate from the adiabatic vacuum \cite{27} during the transition
period from inflation to reheating, it was shown that heavy scalar particles
of this kind can be effectively produced if their mass is of the order of
or less than the mass of an inflaton field.

The method of Hamiltonian diagonalization discussed in Sections 3 and 4
was also applied in Ref.~\cite{89} to describe the creation of conformally
coupled to gravity superheavy particles in the model of quintessential
inflation \cite{90}. It was argued \cite{89} that the subsequent decay of
these particles leads to a formation of the relativistic plasma and
eventually results in the universally accepted picture of the hot Universe.

Similar to the creation of particles in a nonstationary electric field, which
has a condensed matter analogy with quasiparticles in graphene, there are
the condensed matter analogies to the particle creation in cosmology.
Recently it was found \cite{91} that expanding Universe resembles the
ultracold quantum fluid of light, where a spatial coordinate plays the
role of time. According to the authors, they observed the acoustic peaks
in the power spectrum which is in quantitative agreement with theoretical
prediction. The observed spectrum was compared with that of cosmic microwave
background power spectrum. Another possibility of simulation of the process
of particle creation in expanding Universe on the laboratory table by means
of ultra-cold atoms in Raman optical lattices was considered in
Ref.~\cite{92}.

Some more recent publications devoted to the effect of particle creation
in expanding Universe are reflected in the review \cite{93}.

\section{Discussion}

In the foregoing, we have considered the prediction of the Universe
expansion made by Alexander Friedmann a century ago that holds the
greatest importance and interest today. Particular attention has
been given to the way on how this discovery was made. According
to the adduced arguments, it is not an accident that such a
breakthrough result was obtained by a mathematician. Several
outstanding physicists, including great Einstein, worked on the
same subject, but they were tied by some additional considerations
of methodological character implying the static character of our
Universe.

Quite to the contrary, Friedmann restricted himself to
only the necessary minimum assumptions, such as the homogeneity and
isotropy of space, and searched for the formal mathematical
consequences following from the fundamental Einstein equations with
no prejudice. In doing so, Friedmann discovered that
typical cosmological solutions of Einstein equations describe the
expanding Universe. This example shows that the mathematical
formalisms of fundamental physical theories, such as the general theory
of relativity, may in some sense, be more clever than their creators
and again raises a question raised by Wigner about the unreasonable
effectiveness of mathematics in natural sciences \cite{24}.

The importance of Friedmann's prediction of the Universe expansion is
difficult to overestimate. After a comprehensive experimental
confirmation, the concept of expanding Universe laid the groundwork for
the modern picture of the world. As substantiated in Ref.~\cite{94},
this fact gives grounds to include the name of Friedmann along
 with the names of Ptolemy, Copernicus, and Newton who created
the scientific pictures of the Universe accepted in previous epochs.

Expansion of the Universe leads to many outstanding consequences and
one of them, foreshadowed by Erwin Schr\"{o}dinger, is the creation
of particle-antiparticle pairs from the vacuum of quantized fields.
According to a comparison performed in Sections 3 and 4, the effect of
creation of particles in the expanding Universe is mathematically
analogous to that in the nonstationary electric field in spite of
quite different physical situations in both cases.

The main results obtained in the literature on the creation of
particles in expanding Universe by the method of Hamiltonian
diagonalization and other methods show that this effect played an
important role at the very early stages of its evolution and,
especially, during the transition period between the inflationary
and radiation dominated epochs. According to the results obtained,
the effect of pair creation could also contribute to the formation
of dark matter.

\section{Conclusions}

To conclude, in this brief review, devoted to one hundred anniversary
of Alexander Friedmann's prediction of the Universe expansion, we
have considered several facts of his biography which were helpful for
making this outstanding discovery. The results published by
Friedmann in 1922 \cite{4} and 1924 \cite{5} were presented above with
a stress on the role of mathematics in their obtaining. Some historical
facts, including the dispute with Albert Einstein, and further
developments of the Friedmann cosmology are elucidated.

The Universe expansion leads to the quantum creation from vacuum of the
particle-antiparticle pairs. This effect was discussed above in close
connection with a more familiar effect of pair creation by the
nonstationary electric field. The comparison studies of these two effects
by the method of Hamiltonian diagonalization was perfomed and both
the similarities and distinctions between them were analyzed. Several
results for the numbers of scalar and spinor pairs created at different
stages of the Universe evolution are presented. Special attention was paid
to the inflationary stage of the Universe evolution and to the transition
period to the epoch of the radiation dominated Universe where the effect
of particle creation was of primary importance for the formation of
relativistic plasma and cold dark matter.

By and large, the prediction of the Universe expansion made by Alexander
Friedmann laid the foundation for a development of modern cosmology
during the last century and offered possibilities for the description
of vacuum quantum effects in a nonstationary space-time by the
formalism of quantum field theory in the presence of external fields.

\funding{
This work was partially funded by the
Ministry of Science and Higher Education of Russian Federation
("The World-Class Research Center: Advanced Digital Technologies,"
contract No. 075-15-2022-311 dated April 20, 2022).
It was also partially carried out in accordance with the Strategic
Academic Leadership Program "Priority 2030" of the Kazan Federal
University.
}

\acknowledgments{The author is grateful to G.L.Klimchitskaya for helpful 
discussions.}


\begin{adjustwidth}{-\extralength}{0cm}

\reftitle{References}

\end{adjustwidth}
\end{document}